The 7[th] International Conference Interdisciplinarity in Engineering (INTER-ENG 2013)

# Data mining – past, present and future – a typical survey on data streams

## M.S.B. PhridviRaj[a,]*, C.V. GuruRao[b]

*[a]Department of CSE, Kakatiya Institute of Technology and Science, Warangal, INDIA*
*[b] Department of CSE, S.R. Engineering College (Autonomous), Hasanparthy, Warangal,INDIA*

## Abstract

Data Stream Mining is one of the area gaining lot of practical significance and is progressing at a brisk pace with new methods, methodologies and findings in various applications related to medicine, computer science, bioinformatics and stock market prediction, weather forecast, text, audio and video processing to name a few. Data happens to be the key concern in data mining. With the huge online data generated from several sensors, Internet Relay Chats, Twitter, Face book, Online Bank or ATM Transactions, the concept of dynamically changing data is becoming a key challenge, what we call as data streams. In this paper, we give the algorithm for finding frequent patterns from data streams with a case study and identify the research issues in handling data streams.

*Keywords:* Clustering; Streams; Mining; Dimensionality reduction; Text stream; Data streams

## 1. Introduction

Data mining is a process of discovering hidden patterns and information from the existing data. The difference between data in the databases and a data warehouse is in a database the data is in the structured form where as in the data warehouse the data may or may not be present in the structured format. The structure of the data may be defined to make it compatible for processing. Hence in data mining; we also need to primarily concentrate on cleansing the data so as to make it feasible for further processing. The process of cleansing the data is also called as noise elimination or noise reduction or feature elimination. The process of cleansing data can be either made by using tools such as ETL, tools available in the market or may be done by using various suitable techniques available. The important aspect for consideration in data mining is whether the data considered is static or dynamic. Handling static data is comparatively much easier to handling dynamically varying data. In the case of a static dataset, the entire data is available for analysis purpose in hand before processing and is generally not a time varying data. However dynamic data refers to high voluminous continuously varying information which is not a stand still data and also is not at the hand for processing or analyzing.

--------

* Corresponding author. Tel.: +9030076521.
  *E-mail address:* prudviraj.kits@gmail.com

Data mining requires an algorithm or method to analyze the data of interest. Data may be a sequence data, sequential data, time series, temporal, spatio- temporal, audio signal, video signal to name a few. The concept of data streams has gained a lot of practical interest in the field of data mining. A data stream is an infinite sequence of data points defined usually either using time stamps or an index. We may also view data in the data streams as equivalent to a multidimensional vector containing integer, categorical, graphical with the data in structured or unstructured format. If the data is not structured we may have to transform in to a suitable format for processing by the algorithm being used. With the very high voluminous structured or unstructured continuous data being generated from various applications and devices, the concept of data is no more static but is turning out to be dynamic. This brings a lot of challenges in analyzing the data. Traditional data mining algorithms are not suitable for handling data streams because the algorithms designed perform multiple scans over the data which is not possible when handling the data streams. This brings actual challenge before the data mining researchers working in the area of data streams.

Further, Many of the existing data mining algorithms available for clustering, classification and finding frequent pattern in the literature are suitable for only static data sets and are no more practically suitable for handling data streams or for mining the stream data. Data streams may be time series or temporal or spatio temporal. The concept of clustering and classification is widely used and turned out as a choice of typical interest among the current data mining researchers. Section 2 discusses various related works in detail. In Section 3, we discuss various research issues in data mining and problems in handling data streams. We conclude the survey in section 4 finally.

## 2. Related works

In case of data streams, the number of distinct features or items that exist would be so large which makes even the amount of on cache memory or system memory available not suitable for storing the entire stream data. The main problem with data streams is the speed at which the data streams arrive is comparatively much faster than the rate at which the data can be stored and processed.

In the ACM KDD International conference held in 2010, the authors discuss the problem of finding the top-k frequent items in a data stream with flexible sliding widows [3]. The idea is to mine only the top-k frequent items instead of reporting all the frequent items. But the crucial factor or limitation that evolves here is the amount of memory that is required still for mining w.r.t to finding of top-k frequent items is still a bounding factor. The authors finally discusses that there exists however a memory efficient algorithms by making some assumptions.

In [2] the authors focus on developing a framework for classifying dynamically evolving data streams by considering the training and test streams for dynamic classification of datasets. The objective is to develop a classification system in which a training system can adapt to quick changes of the underlying data stream.

The amount of memory available for mining stream data using one pass algorithms is very less and hence there is chance for data loss. Also it is not possible to mine the data online as and when it appears because of mismatch in speed and several other significant factors.

In [4] the authors discuss the method of finding most frequent items by using a hash based approach. The idea is to use say 'h' hash functions and build the hash table by using linear congruencies. Data streams can be classified into two types as 1. Offline data streams and 2. Online data streams.

In [6] the method of singular valued decomposition is used to find the correlation between multiple streams. The concept of SVD was particularly used to find offline data streams. Clustering text data streams is one of the topics which have evolved as important challenge for data mining researchers. The problem of spam detection, email filtering, clustering customer behaviours, topic detection and identification, document clustering are a few of typical interest to data mining researchers.

In [7], Liu et.al discuss on clustering text data streams. The idea is to extend the existing semantic smoothing model which works well with static data streams for clustering dynamic data streams. The authors propose two online clustering algorithms OCTS and OCTSM for clustering massive text data streams.

A tremendous amount of data is generated from web every instant in various forms such as social networks, data from sensors, face book and twitter. The emerging data from web also called as Text message stream which is generated from various instant message applications and internet relay chat.

This has become a prime topic which has become a hot topic of interest to the researchers working in the area of data mining and has a lot of scope to work to be contributed by the research community.

In ACM SIGIR held in 2006 the authors, Shen, Yang et.al [8], propose the method of detecting the threads in dynamic data streams. The paper discusses three variations of single pass clustering algorithm followed by a novel clustering algorithms based on linguistic features. A method of reducing the dimensionality of streaming data using a scalable supervised algorithm is proposed in [9].

The limitations of PCA, LDA and MMC approaches are discussed. The authors point out the unsuitability of MMC for streaming data. A supervised incremental dimensionality reduction algorithm is proposed to meet the requirements of streaming data set. In [10] the authors show that the most cited result hoeffdings bound is invalid

## 3. Research issues in handling data streams

The Table.1 below shows comparison of processing data in database and data streams

Table 1. Processing data vs data streams.

| No | Parameter | Database | Data streams |
|----|-----------|----------|--------------|
| 1 | Data access | May or may not be Sequential | Sequential |
| 2 | Available Memory | Flexible | Limited memory |
| 3 | Data spread | Need not be | Distributed |
| 4 | Computation Results | Accurate | Approximate findings |
| 5 | Data Scan | Flexible | Limited to one scan |
| 6 | Algorithms | Processing time is not a constraint | Processing time is most important as data may skip |
| 7 | Sampling | Not required | Complex to decide when to sample data |
| 8 | Data speed | Can be ignored | Data arrival rate is higher than processing rate |
| 9 | Data modelling | Persistent | Modeled as Transient Data streams |
| 10 | Data Schema | Static | Dynamic |

### 3.1. Memory constraint

The primary factor which we need to narrow down to, when handling streaming data is the amount of memory required by the stream mining algorithm used or being designed. The algorithms designed for static data also called non-stream data are no more feasible for handling streaming data or data streams. The challenge in handling data streams is that the data being generated is not regular and mostly generated at irregular time intervals. The focus of the algorithm must essentially be made on optimizing the memory utilized by the algorithm for processing. Since the data in the data stream is dynamic, it is not in the hand at the instant of processing it and also it may not be in proper format. The amount of data generated is typically very huge and keeps growing and adding as time progresses and this makes the situation more complicated when handling the streaming data.

*3.2. Data preprocessing*

The data processing task is also one of the criteria which must be taken care in the process of data mining. The data input to a data mining algorithm need not be in proper format and is hence not suitable for processing efficiently. In such a case, we need to see the data is in proper format so that it is suitable for processing. This case generally arrives when we try to mine the data using the existing data mining tools or algorithms. Different Data mining tools available in the market have different formats for input which makes the user forced to transform the existing input dataset into the new format. This itself is very time consuming, laborious and has a chance of data loss as the data is to be entered manually into a new format to be supported by the tool.

The second factor is the dimensionality of the data. The dimensionality of the data can be very high which makes it much more complex for analysis. Also much of the processing time of the algorithm is killed in this case. Proper care has to be taken care in reducing the dimensionality of the data which otherwise can be very destructive w.r.t to the algorithms running time. For example, if we consider the text data, there shall be several unnecessary, unwanted features, do not contribute significantly to the decision making or analysing.

Reducing dimension of the input data set also helps in reducing the memory requirement and thus achieves space efficiency of the data mining algorithm. Several geometric and statistical techniques are being used by researchers in this area but this is still a place where data mining researchers are getting stuck.

*3.3. Challenges in designing frequent pattern mining algorithms*

Though much amount of memory is available currently these days, the memory factor is dominating the field of data mining and posing challenges for the researchers working in the field of data mining, more specifically in the field of data streams. The reason for this is the requirement of main memory which is usually very less as compared to the other memory available outside the processor. Designing frequent pattern mining algorithm for handling data streams is quite challenging as we have the main memory limitation. Finding frequent item using single pass algorithms is also not practical in case of handling data streams because of the dynamic nature of data streams.

*3.4. Choice of data structure*

The choice of a suitable and effective data structure is also one of the criteria that need to be taken care in design of algorithms for handling data streams.

*3.5. Identifying data distributions and target concepts*

Another important factor when consider mining data streams is to identify the changes in the data distributions and target concepts over time. Identifying or exploring theses variations and adapting suitable classifiers to concept drifts is also one of the biggest challenges before data mining researchers in designing new scalable algorithms for handling data streams.

*3.6. Dimensionality reduction*

Though the mathematical or statistical techniques were available in the literature earlier, the importance and suitability of these techniques is being mostly explored by data mining researchers in the very recent times. The problem of dimensionality reduction is studied by applying the mathematical and statistical approaches. However, handling dimensionality is still a problem in data mining when handling data with and without streams.

## 4. Frequent patters from data streams

We consider the method of finding frequent items from data streams using sliding windows. Let S be the sliding window of Size n with I = $\{i_1, i_2, i_3 \ldots i_m\}$ as the set of all available items. Depending on the type of the transaction done, a transaction $T_j$ may contain the entire item set denoted as I or only a proper subset of I as its items.

Let A and B be any two items of a transaction. The binary vector for A and B are denoted by Bin-Vector (A) and Bin-vector (B).

Now, as the size of the sliding window is restricted to a count of n transactions, we restrict the representation of the binary vectors for items A and B to be of the form n-bit binary vector as

Bin_Vector(A) = A1 A2 A3 A4…. An and

Bin_Vector(B) = B1 B2 B3 B4…. Bn.                 where n is the size of sliding window.

If an item A is present in the transaction Ti then the corresponding bit of the Bin-vector(A) is set to 1. Similarly, if the item A is not present in a given transaction $T_i$ the corresponding $i^{th}$ bit of the Bin-vector (A) is set to 0. This is shown as the first level nodes of the Frequent-Pattern-Generation-Tree called FPGT in fig.1.

**Definition.1.** Let Bit-1 and Bit-2 be two one bit ternary numbers. We define a function F over Bit-1 and Bit-2 as given in Table. 2

Table 2. Definition of the function F.

| Bit -1 | Bit-2 | F(Bit-1,Bit-2) |
|--------|-------|----------------|
| 0 | 0 | Z |
| 0 | 1 | 0 |
| 1 | 0 | 0 |
| 1` | 1 | 1 |
| 0 | U | Z |
| U | 0 | Z |
| U | U | Z |
| 1 | U | Z |
| U | 1 | Z |

**Definition-2:** The ternary function F in the Table. 2 takes input as two 1-bit ternary numbers and outputs 1-bit value as 0 or 1or Z. We extend the function F in Table.1 to compute over two ternary feature vector of itemset by applying F for each corresponding bits of the feature vectors.

We assume the itemset to be static. However if the itemset is dynamic and gets added later we just need to generate a new link from the root node. A Node in the FPGT tree consists of three fields

1. First field represents the Item ID or name
2. Second field is the Binary Feature Vector representation of item
3. Third field indicates the count of 0's in the Binary Feature Vector.
4. Status of node denoting Live or Dead.

The main problem in handling data streams is memory constraint because we are restricted to a single scan of the database. The algorithm defined below performs only one time scan of the database initially and uses the information to find frequent patterns using frequent pattern generation tree.

*4.1. Algorithm for Frequent_Pattern_Generation_ Tree (Itemset, Frequent patterns)*

Let A be the any item, S be the sliding window, Si is the $i^{th}$ sliding window and T = {$T_1$, $T_2$, $T_3$,……….$T_n$} be the transactions in current the sliding window

**Step.1:** Start with the root node and generate a node for each item in the list of transactions of the sliding window.

This is the first node in the frequent pattern generation tree which we call as start node or root node consisting of m fields where m is the number of items. The fields of the root node contain no information but are just links pointing to m items of the transaction data set.

We perform the data scan of the entire database for the first time and store which item belongs to which transaction.

**Step2: // Compute m-itemsets with i=2, 3, 4….m**

For each item in the FPGT tree generated in Step1

For each of its corresponding siblings towards its right

// We call it as ternary feature vector because we have three values 0, 1, U and quaternary as we each node has 4 fields

Create a new quaternary node with four fields with the first field as 2-item set name, second field as n-bit ternary feature vector and third field containing count of 1s in ternary feature vector and last field indicating status of node as live or killed.

If (Support_value (2-itemset) of E-node generated < user_threshold)
kill the corresponding node of the tree and mark it as dead node
else
retain the node and mark the node as live node

**Step3:**

For each node generated in step-2 of the partial FPGT tree

Consider only parent nodes of the node at level-i+1.
If there exists an node with itemset in level-i which is subset of current E-node at level i+1 and has same support value
Kill the node at level-i. // This is because node at level-i is a subset and has no impact on the deletion.
Else Retain the node for future

**Step-4:** Repeat step-3 till we get no new node is generated.

**Step-5:** Display the nodes with first k-larger values which form top-k frequent items.

End of the algorithm

*4.2. Case Study*

Let the incoming transaction flow data stream is as shown below in Table. 3.

Table.3 Data stream with seven transactions and sliding window of size=5.

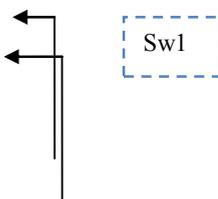

| Transaction | Items |
|---|---|
| T1 | A  B  C |
| T2 | B  C  D |

| T3 | A | B | C | |
|---|---|---|---|---|
| T4 | | B | C | |
| T5 | | | B | | D |
| T6 | A | B | C | D |
| T7 | | | | C | D |

Sw2

Initially form the binary vectors for all the static list of items in itemset defined as I= {$I_1$, $I_2$, $I_3$, $I_4$} = {A, B, C, D} as

Bin-Vector (A) = 10100
Bin-Vector (B) = 11111
Bin-Vector (C) = 11110
Bin-Vector (D) = 01001

Where A, B, C, D are all 5-bit Binary Bit vectors.
Consider A=10100. This represents that item A is present in transactions 1 and 3 only. In Binary vector, 1
Indicates presence and 0 indicates absence. Let user defined threshold is 20%. This means support, S=20%. In other words as there are 5 transactions, item must be present in at least one transaction.

This means that

Support (A) = 2
Support (B) = 5
Support (C) = 4
Support (D) = 2

Now since all items have the count of 1s more than 1. This means all nodes are live as shown in Level-1 of the Tree. The following figures are self-explanatory and shows the trace of the proposed algorithm

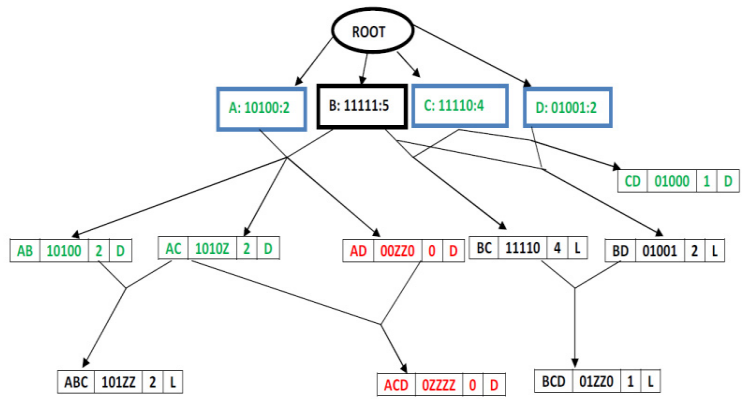

Fig.1. Frequent Pattern Generation Tree showing creation of root node and handling SW1

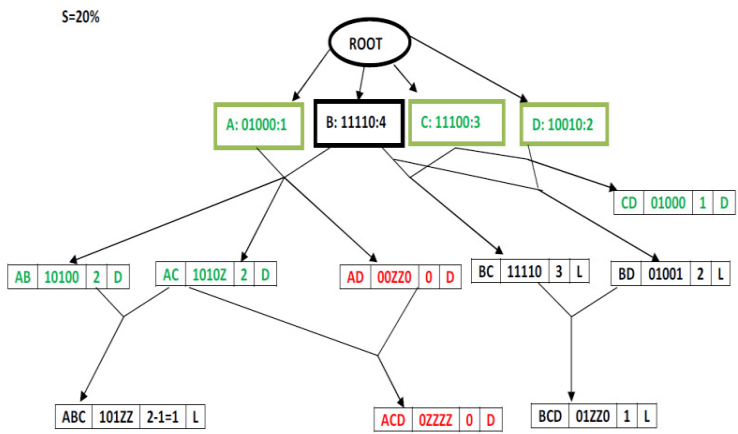

Fig.2. Frequent Pattern Generation Tree showing deletion of nodes when SW1 slides and SW2 is active

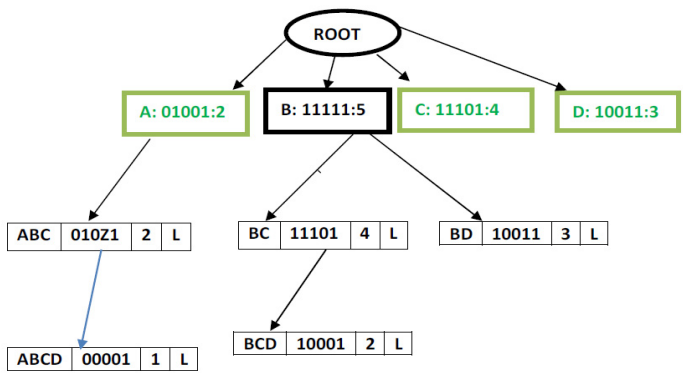

Fig.3. Final Frequent Pattern Generation Tree

In Fig.1, Fig.2, Fig.3 the nodes labeled black are live nodes and active. The nodes labeled green are not closed nodes or items. The nodes labeled red are killed nodes. The nodes at the first level are never killed but just labeled green to indicate that they are not closed nodes.

Once we have all the frequent items (live nodes), we can find the top-k frequent patterns by displaying the first K-items from higher support values to lower support values.

If K=3 then the top-3 frequent terms are {B, BC,BD}.

## 5. Conclusion

Due to unlimited, enormous, high volume data getting generated from various applications as data streams it is quite typical to handle them because of their dynamic, irregular and variant nature. The problem of handling streams for clustering, classification and topic detection is still a challenge and has a wide chance of exploration for data mining researchers to carry their work. In this paper, we find the frequent pattern from data streams using the algorithm defined which uses frequent pattern generation tree. The workflow of the algorithm is traced. The algorithm lists all possible frequent patterns and can be used to find top-k frequent items.